\def\GeV{{\rm GeV}}

\def\GeVc{{\rm GeV/c}}
\def\fm{{\rm fm}}
\def\etal{{\it et al.}}

\def\FullStop{.}

\def\JournalRef#1#2#3#4{{\em #1}\/ {\bf #2} (#3) #4}

\def\PhysRev{Phys.\ Rev.}
\def\PRL{Phys.\ Rev.\ Lett.}
\def\NuclPhys{Nucl.\ Phys.}
\def\PhysLett{Phys.\ Lett.}
\def\ZPhys{Z.\ Phys.}

\def\LAT#1#2#3{\count255=#1\advance\count255 by 1
\JournalRef{\NuclPhys}{B (Proc. Suppl.) #2}{19{\the\count255}}{#3}}

\def\EqnRef#1{Equation~\ref{#1}}
\def\FigRef#1{Figure~\ref{#1}}
\def\TabRef#1{Table~\ref{#1}}
\def\SectRef#1{Section~\ref{#1}}

\def\BorderBox#1#2{\vbox{\hrule height #1%
                         \hbox{\vrule width #1%
                               #2%
                               \vrule width #1%
                              }%
                         \hrule height #1%
                        }%
                  }

\def\BraKet#1#2#3{\left<#1\left|#2\right|#3\right>}

\def\BiLinear#1#2#3#4#5{\overline{#1}#2\,#3\,#4#5}

\def\simle{\mathrel{\rlap{\raise 0.511ex\hbox{$<$}}%
                         {\lower 0.511ex\hbox{$\sim$}}%
                }}

\def\simge{\mathrel{\rlap{\raise 0.511ex\hbox{$>$}}%
                         {\lower 0.511ex\hbox{$\sim$}}%
                }}

\def\TextStyleOver#1#2{%
        {\kern.05em\raise.3ex\hbox{\the\scriptfont0 $#1$}%
        \kern-.1em{\the\scriptfont0 /}%
        \kern-.1em\lower.2ex\hbox{\the\scriptfont0 $#2$}}}

\def\Err(#1){\pm #1}
\def\aErr(+#1-#2){\left(\vphantom{0}^{+#1}_{-#2}\right)}

\documentstyle[twoside,epsf,fleqn,espcrc2]{article}

\newdimen\bls
\bls=\baselineskip
\advance\bls -1ex

\title{
\vskip-2.0em
\vbox to 0pt {\vss \hbox to 6.25truein {\hfil {\small FERMILAB-CONF-96/017-T}}}
\vskip1.0em
Semileptonic Decays: an Update Down Under%
\thanks{Plenary talk presented at Lattice~'95 Melbourne, Australia.}}

\author{James N.\ Simone \\ [\bls]
	Theory Group, Fermilab, P.O. Box 500, Batavia, IL, 60510, USA
	}

\begin{document}

\begin{abstract}
Heavy-meson semileptonic    decays calculations   on the lattice   are
reviewed.    The   focus    is    upon  obtaining   reliable    matrix
elements.  Errors that  depend upon  the lattice  spacing, $a$, are an
important  source of  systematic  error.   Full $O(a)$ improvement  of
matrix elements for arbitrary-mass four-component quarks is discussed.
With  improvement,  bottom-quark  matrix elements  can  be  calculated
directly using  current  lattices.   Momentum  dependent  errors   for
$O(a)$-improved  quarks and statistical noise  limit momenta to around
$1\,\GeVc$  with current lattices.   Hence, maximum recoil momenta can
be reached for $D$ decays while only a fraction  of the maximum recoil
momentum can  be reliably studied for  the light-meson  decay modes of
the $B$.    Differential    decay  rates   and partial     widths  are
phenomenologically important quantities  in  $B$  decays that can   be
reliably determined with present lattices.
\end{abstract}

\maketitle

\section{INTRODUCTION}\label{sect:Intro}

\begin{figure}
\vskip-0.8cm
\BorderBox{0pt}{
\epsfxsize=3.0in \epsfbox[60 360 600 720]{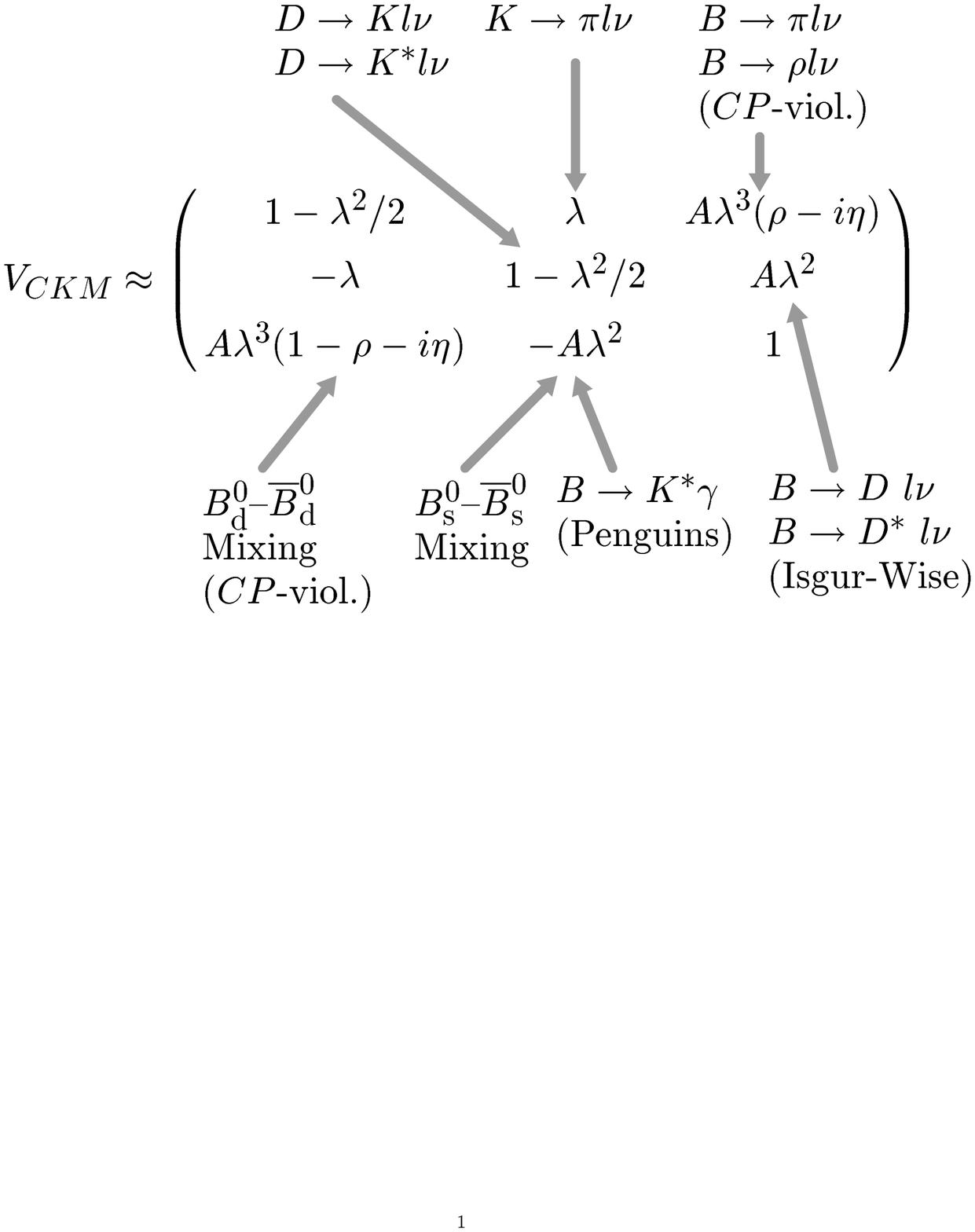}
}
\vskip-1.0cm
\caption{%
The  CKM matrix   in  the Wolfenstein  parameterization.   Decay  rate
measurements  yield  CKM   matrix elements provided   hadronic  matrix
elements are known.}
\label{fig:CKM}
\end{figure}

Semileptonic  decays provide much  needed information about CKM matrix
elements involving heavy quarks.  Precise CKM values are important for
finding the origin of $CP$ violation and  revealing new physics beyond
the Standard Model.  There are many avenues to the CKM matrix elements
open to  lattice studies\cite{LAT95reviews}.   \FigRef{fig:CKM}  shows
the  Wolfenstein parameterization of the  CKM  matrix with some decays
studied on the  lattice.   $B$ semileptonic decays are  of  particular
interest as ways of determining $|V_{ub}|$ and $|V_{cb}|$.

\begin{figure}
\BorderBox{0pt}{
\epsfxsize=2.9in \epsfbox[70 324 480 580]{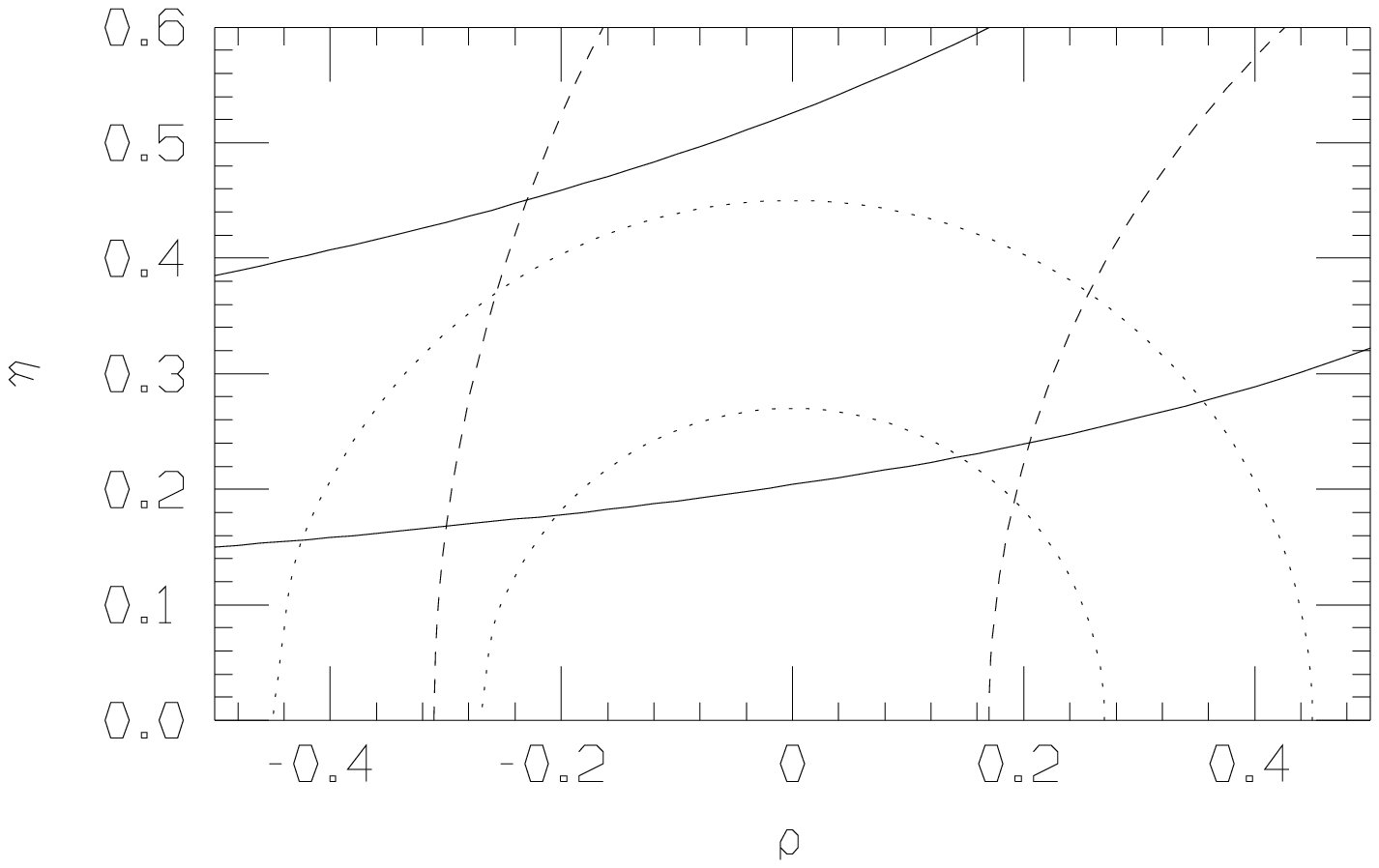}
}
\vskip-1.0cm
\caption{%
The $(\rho,\eta)$ plane.  The  semicircular allowed region centered at
$(0,0)$ (dotted lines) is from inclusive $B$ semileptonic decays.  The
region between   dashed  lines is  allowed  by  $B^0$-$\overline{B}^0$
mixing.  The region    between solid lines comes  from  $CP$-violating
$K^0$-$\overline{K}^0$ mixing. Rosner\protect\cite{RosnerDPF}. }
\label{fig:RhoEta}
\end{figure}

The values of $|V_{ub}|$ and $|V_{cb}|$ from semileptonic decays place
important  constraints upon Wolfenstein  parameters $\rho$ and $\eta$.
From    \FigRef{fig:CKM}  the      ratio    of   CKM   elements     is
$\lambda\sqrt{\rho^2+\eta^2}$  where  $\lambda$ is   the sine   of the
Cabbibo  angle. The  ratio   scaled   by $\lambda$  then defines     a
semi-circle centered at  $(0,0)$      in the $(\rho,\eta)$      plane.
Uncertainties in the   values of the  CKM  matrix  elements widen  the
semicircle into the allowed region shown between the dotted arcs in
\FigRef{fig:RhoEta}.

The figure also shows allowed regions in  the $(\rho,\eta)$ plane from
$B^0$-$\overline{B}^0$ mixing and $K^0$-$\overline{K}^0$ mixing.  With
more  precise measurements  of   the CKM matrix  elements  the allowed
regions  will shrink.   If all  the  regions no longer intersect  this
would   indicate   an    inconsistency with     the  Standard   Model.
Uncertainties in hadronic  matrix elements are now  one of the largest
sources of uncertainty for CKM matrix elements.

The ratio $|V_{ub}|/|V_{cb}|$ is now best determined by looking at the
lepton energy spectrum for  inclusive decays  in the end-point  region
above the charm production limit\cite{CLEOincl,ARGUSincl}.  This ratio
depends upon   theoretical form factors.   Using current   form factor
models,   the       Particle       Data  Group\cite{CKMPDG94}   quotes
$|V_{ub}|/|V_{cb}|=0.08\pm0.02$   where  the 25\%   error  is   due to
combined  statistical    and theoretical  uncertainties.  Three-family
unitarity  of the CKM  matrix and constraints  from  other CKM element
measurements  yield  $|V_{ub}|$  values that   range  from $0.002$  to
$0.005$ at the 90\% confidence  level.  Thus, $|V_{ub}|$ is only known
to within a factor of two.

The  CLEO II collaboration has  recently  reported the first exclusive
measurements      for       $B$     light-meson   semileptonic   decay
modes\cite{Gibbons95}.   These measurements  promise better methods of
determining  $|V_{ub}|$.  Exclusive  branching  ratios  determinations
also require form factors as  theoretical inputs.  With current models
the CLEO results have an  uncomfortably large theoretical uncertainty.
Hence, there is  an   essential  need  for reliable hadronic    matrix
elements.    Given  the   phenomenological  importance   of  $V_{ub}$,
light-meson decays modes  for the $B$  meson deserve high  priority in
lattice studies.

The  motivation for using  the lattice is clear:  it provides a direct
numerical  solution  of  QCD.  At  present, the  lattice   is the only
systematically improvable, nonperturbative calculation method for QCD.
Lattice   technology for    semileptonic   decays   is  already   well
established\cite{SLreview}.  Systematic  errors  must be understood to
obtain  reliable results from the   lattice. Lattice discreteness is a
source of systematic  errors.  Discretization errors for heavy  quarks
and momentum dependent  errors are two  important potential sources of
error considered in \SectRef{sect:WilsonQuarks}.

Full  $O(a)$   improvement    for  on-shell  matrix   elements    with
arbitrary-mass   four-component quarks is  discussed.  Improved matrix
elements for  bottom quarks  can   be  studied on  present   lattices.
Improvement increases  the  reliability of  lattice  calculations  and
allows coarser lattices to  be used.  Computations then  become faster
and  cheaper.   Less  costly   numerical  calculations  make   it cost
effective to do systematic studies of other sources of error.

\SectRef{ssect:MomentumErrors}    discusses     momentum     dependent
errors.   Estimates  for momentum dependent  errors in $O(a)$-improved
matrix elements on present lattices show that errors are probably less
than  20\% for momenta below  $1\,\GeVc$.  Reducing these errors would
require smaller lattice  spacings, adding $O(a^2)$ improvements to the
quark  action  or extrapolations   to zero  lattice  spacing at  fixed
physical momentum.

\SectRef{sect:Ddecays}  is    a   survey  of   $D$    decays   on  the
lattice. Maximum meson recoil momenta in the  $D$ rest frame are below
$1\,\GeVc$ for  semileptonic  decays.  Momentum dependent  errors  are
then probably  below 20\% in present  studies.  Hence, form factors at
maximum recoil   momentum ($q^2=0$) can be  extracted.    A summary of
lattice $D\to  K$ and $D\to K^*$ form  factors however, shows no trend
with  lattice spacing.   Therefore, high precision   tests for lattice
spacing dependence are still needed.

Light-meson decay    modes    of   the   $B$   are     considered   in
\SectRef{sect:BuDecays}.  On  present  lattices,  one approach  to $B$
decays is to study $D$ decays and then to extrapolate, guided by heavy
quark  symmetry, to the $B$.    Heavy-quark scaling of hadronic matrix
elements is tested by calculating  $O(a)$-improved matrix elements for
charm, bottom and static  quarks.  Matrix elements vary smoothly  with
quark mass.  Therefore, when heavy quark  symmetry is used to find $B$
matrix elements,  interpolations including static  matrix elements are
apt  to  be more reliable  than  extrapolations based on charm results
alone.

Light mesons  have a  maximum  recoil momentum of  $2.6\,\GeVc$ in $B$
decays.   Present  $B$ calculations  are   then restricted to  only  a
fraction  of  the full momentum range.    Predictions  at large recoil
momenta are then potentially unreliable.  Differential decay rates and
partial widths are suggested as useful quantities that can be reliably
calculated with present lattices.

\SectRef{sect:BcDecays} reports  upon decays  with  heavy hadron final
states.  Studies have focused upon mesons in order to understand $B\to
D$  decays and extract  $|V_{cb}|$.  Baryon decays  are now also under
study.   Baryons  offer a new   laboratory  for heavy  quark  symmetry
studies   --   including baryon   Isgur-Wise  universal  form factors.
Finally, two  studies of Heavy Quark Effective  Theory  on the lattice
are discussed.  One study examines nonperturbative renormalization for
lattice  Heavy Quark Effective Theory  while the second demonstrates a
dramatic   reduction   in statistical  noise  when    using  optimized
wavefunctions in calculations of the meson Isgur-Wise function.

\section{ARBITRARY-MASS QUARKS}\label{sect:WilsonQuarks}

A quark is considered heavy  when $m_0\gg\Lambda_{QCD}$ as is the case
for      charm         and       bottom   quarks.      Wilson      and
She\-ik\-hole\-slami-Woh\-lert (SW)  heavy quarks  are widely  used in
matrix  element studies.  On  present  lattices heavy-quark masses are
typically  a large fraction  of   the lattice cutoff   $m_0\simge1/a$.
Errors  arising   from the discrete    lattice can then  be important.
Standard Wilson quarks  have $O(a)$ errors  in matrix elements.  Since
decreasing the lattice spacing so that $am_0\ll 1$  for a bottom quark
is  extremely costly, improvements  which remove discretization errors
on present lattices become   an attractive alternative.   This section
focuses on a method for systematically removing lattice-spacing errors
in four-component quarks of arbitrary mass.

\begin{figure}
\BorderBox{0pt}{
\epsfxsize=3.0in \epsfbox[124 82 480 374]{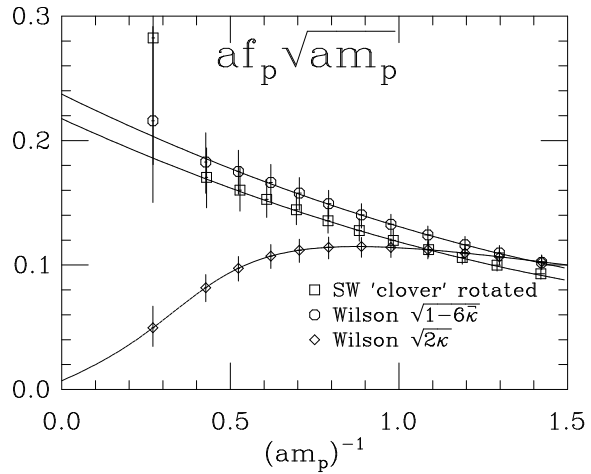}
}
\vskip-1.0cm
\label{fig:fsqrtM}
\caption{%
The quantity $af_p\protect\sqrt{am_p}$ for  heavy mesons.  Three types
of    quarks are  compared:  conventional   Wilson quarks  (diamonds),
improved  Wilson  quarks (octagons) and  Clover  quarks (squares).  At
large mass  conventional   Wilson quarks  have  large   discretization
errors.   The $D$  meson  mass  is $(a   m_D)^{-1}\approx 1$ on   this
lattice. }
\end{figure}

Lattice-spacing errors  can  be observed  in results for  pseudoscalar
decay constants.  \FigRef{fig:fsqrtM}   shows decay constants  from  a
study using both Wilson and SW quarks on a lattice where the $D$-meson
mass is  $am_D\sim1$\cite{SimoneLAT92}.   Wilson results are indicated
by  the diamond  symbols in the  figure.   Conventional  Wilson quarks
incorrectly lead to $af\sqrt{am}=0$ in the infinite-mass limit.  Hence
discretization errors  overwhelm Wilson quarks  at  large masses.  The
figure shows that extrapolations of Wilson results from the $D$ to the
$B$ are apt to be unreliable.

Errors  in on-shell matrix elements  can  be reduced systematically by
improving   the   lattice  action    and  currents.   $O(a)$  (Clover)
improvement   for     light   quarks is         based  upon the     SW
action\cite{Heatlie91}.   The  square symbols  in  \FigRef{fig:fsqrtM}
denote Clover decay    constants.  Clover results   extrapolate  to an
infinite-mass      limit     closer    to   the        static    value
$af\sqrt{am}\approx0.32$  indicating    a  sizable  reduction       in
discretization errors at large mass compared to Wilson quarks.

Consider  again the  Wilson results.  Large  discretization errors are
apparently  removed by     adopting  an unconventional    wavefunction
renormalization    for   Wilson   quarks\cite{Mackenzie92,Kronfeld92}.
Results for reinterpreted Wilson quarks (circles in
\FigRef{fig:fsqrtM})  also   have  an infinite-mass  limit   in better
agreement  with    the    static result.  With     the  unconventional
wavefunction renormalization  the  correct free-field quark propagator
is recovered at infinite-mass.  Since improvement  is possible in both
heavy  and  light limits,  the Fermilab  group  then  investigated the
possibility of improvements for quarks of arbitrary mass.

The      Fermilab  program    improves  arbitrary-mass  four-component
quarks\cite{MassiveFermion95}.   It    is equivalent  to    the  usual
improvements  for   light quarks and  to  the  static approximation at
infinite mass. Fully $O(a)$-improved quarks  are obtained using the SW
action and improved  currents.  Bottom quarks can  be  put on existing
lattices without $O(a)$ errors.  Heavy  Wilson and SW quarks are  used
successfully to  study  -onia\cite{ElKhadra94}.  In  $B$-decay studies
uncertainties due   to extrapolations in the   heavy-quark mass can be
eliminated and calculations become   cheaper and faster since  present
lattices can be  used.  Computational   savings can  then be used   to
remove  remaining discretization  errors by repeating  calculations at
several lattice spacings.

\subsection{The Action}\label{ssect:Action}

The $O(a)$  quark action in the Fermilab  formalism is  a sum of three
terms:
\begin{equation}
S_f = S_0 + S_B + S_E   \FullStop
\label{eqn:FermionAction}
\end{equation}
Term  $S_0$  has    dimension-three   and   -four operators   and    a
dimension-five Wilson operator to eliminate doublers:
\begin{eqnarray}
\nonumber
S_0 &=  &m_0 \int\BiLinear{\Psi}{^\prime}{\;}{\Psi}{^\prime} +
\int\BiLinear{\Psi}{^\prime}{\frac{1}{2}(1+\gamma_0)D_0^-}{\Psi}{^\prime} \\
\nonumber
 & &\!\!\!\!\!\!-\int\BiLinear{\Psi}{^\prime}{\frac{1}{2} (1-\gamma_0)D_0^+}{\Psi}{^\prime} +
\zeta\int\BiLinear{\Psi}{^\prime}{\vec{\gamma}\cdot\vec{D}}{\Psi}{^\prime} \\
& &-\frac{1}{2}a r_s \zeta \int\BiLinear{\Psi}{^\prime}{\Delta^{(2)}_s}{\Psi}{^\prime} \FullStop
\label{eqn:Szero}
\end{eqnarray}
Operators  $D_\mu^\pm$ are the forward  ($+$) and backward ($-$) first
differences, $D_\mu$ is the central difference and $\Delta^{(2)}_s$ is
the spatial laplacian.  By setting  $r_s=1$ and $\zeta=1$ action $S_0$
becomes the Wilson action.  The action is asymmetric in space and time
when $\zeta\neq1$.

The chromomagnetic and chromoelectric terms are:
\begin{eqnarray}
S_B &= &-\frac{i}{2} a c_B \zeta \int\BiLinear{\Psi}{^\prime}{\vec{\Sigma}\cdot\vec{B}}{\Psi}{^\prime}
\label{eqn:SB} \\
S_E &= &-\frac{1}{2} a c_E \zeta \int\BiLinear{\Psi}{^\prime}{\vec{\alpha}\cdot\vec{E}}{\Psi}{^\prime} \FullStop
\label{eqn:SE}
\end{eqnarray}
In contrast to the light-quark action where $c_B=c_E$ and $S_B+S_E$ is
``clover'' term, here $c_B$ and $c_E$ may be unequal.

Parameters $r_s$ $\zeta$,  $c_B$    and $c_E$  are functions  of   the
coupling; tree-level values are known. Higher-order corrections can be
obtained perturbatively. These parameters are  also functions of quark
mass.  The mass dependence is treated to all orders.

Values $r_s=1$,   $\zeta=1$ and $c  = c_B  = c_E$ are  used  for light
quarks.  With  $c=1$ the action is the  tree-level  SW action; tadpole
improvement increases  $c$ to  between   $1.4$  and $1.5$  on  typical
lattices.     The    tadpole-improved SW   action    leads   to  fully
$O(a)$-improved light  quarks.  $O(a)$  improvement is  still obtained
for heavy quarks provided currents and  the physical mass are suitably
defined.

\subsection{Physical Heavy-Quark Masses}\label{ssect:QuarkMasses}

In the non-relativistic limit, the Hamiltonian corresponding to $S_f$ is
\begin{eqnarray}
H&\simeq
&\BiLinear{\Psi}{}
{\left(
M_1
+\gamma_0 A_0
-\frac{\vec{D}^2}{2M_2}
-\frac{i\vec{\Sigma}\cdot\vec{B}}{2M_B}
\right.}
{\phantom{\Psi}}{\phantom{^\prime}} \nonumber \\
&&\qquad\phantom{\Psi^\prime}\,
\left.-\gamma_0\frac{[\vec{\gamma}\cdot\vec{D},\vec{\gamma}\cdot\vec{E}]}{8M_E^2}\right)\,
\Psi
\label{eqn:Hamiltonian}
\end{eqnarray}
where $\Psi$ is the  physical field defined in \EqnRef{eqn:NRotation}.
This looks like the Pauli Hamiltonian; $M_1$,  $M_2$, $M_B$, and $M_E$
however are   functions    of $m_0$.  This   Hamiltonian   describes a
non-relativistic quark when  $M_2=M_B=M_E$.   For both Wilson and   SW
quarks, when  $am_0\simge  1$ the  kinetic mass, $M_2$,   and the rest
mass, $M_1$, are significantly  different.  Since $M_1$ is unimportant
for  non-relativistic quarks,   $M_2$ should  then  be considered  the
physical mass.

The  Hamiltonian is  a    useful   guide for estimating   errors   for
non-relativistic quarks.  Systematic  errors  will arise if $M_B$  and
$M_E$ do  not equal $M_2$.   At tree-level, when $am_0\simge 1$ Wilson
quarks  have     $M_B>M_2$ while SW     quarks   have $M_B\simeq M_2$.
Therefore, from   \EqnRef{eqn:Hamiltonian},   Wilson  quarks   can  be
expected to correctly reproduce only spin-averaged energy levels while
SW quarks should also approximate spin-dependent features.

\begin{figure}
\BorderBox{0pt}{
\epsfxsize=3.0in \epsfbox[110 85 500 374]{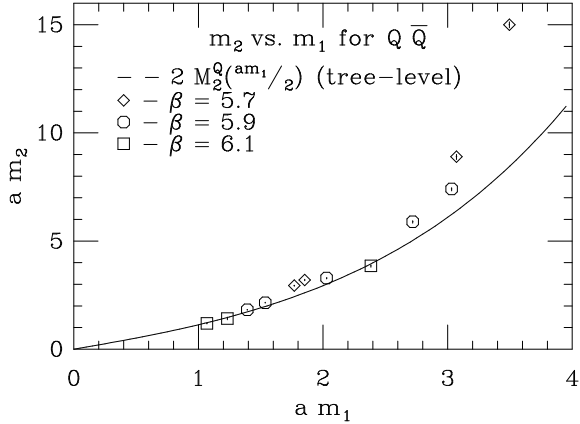}
}
\vskip-1.0cm
\caption{%
Meson kinetic mass, $m_2$,  as a function of the  rest mass,  $m_1$ in
-onia.  Points are lattice  determinations using the energy dispersion
relation.  The  curve is based upon  the tree-level expression for the
quark kinetic mass in terms of the rest mass.}
\label{fig:M2vsM1}
\end{figure}

Physical quark masses can be found non-perturbatively in -onia by using
the non-relativistic energy dispersion relation
\begin{equation}
E_1^{NR}=m_1+\frac{\vec{p}^2}{2m_2}+\ldots \qquad\FullStop
\label{eqn:NREdispersion}
\end{equation}
The  meson  kinetic mass,  $m_2$, and  the   rest mass,  $m_1$, differ
significantly when $am_0\simge1$ for  Wilson and SW quarks.  Charm and
bottom quark  masses are determined by matching  the physical mass and
$m_2$.   Spin-averaged levels are  used;  the physical lattice spacing
may be found using the $1P$-$1S$ mass splitting.

\FigRef{fig:M2vsM1} shows $am_2$ and  $am_1$ values obtained for three
lattice spacings using SW quarks.   Numerical results agree well  with
the curve\cite{CollinsThesis} which  is    based on  the    tree-level
functional relation between quark masses $M_2$ and $M_1$.  Differences
between numerical results and tree-level predictions are, in part, due
to higher-order corrections.  One-loop corrections to quark masses are
being computed\cite{Kronfeld93}.  The figure shows  that the rest mass
systematically underestimates   the physical  mass.    At $\beta=6.1$,
$M_2/M_1$ is about $1.2$ for the charm quark  and $2.8$ for the bottom
quark. Using the rest mass to  determine the physical mass rather than
the   more physical kinetic mass   introduces a  systematic error into
heavy-quark calculations.

\subsection{Currents}\label{ssect:Currents}

\begin{figure}
\BorderBox{0pt}{
\epsfxsize=3.0in \epsfbox[82 77 505 316]{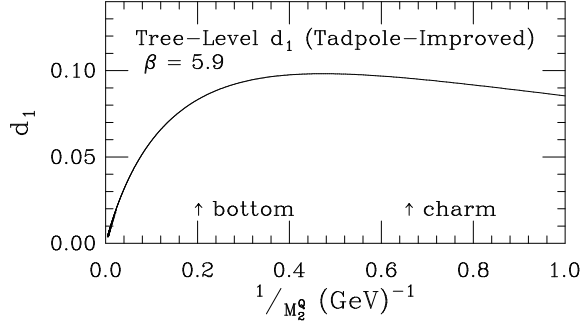}
}
\vskip-1.0cm
\caption{%
Tadpole-improved tree-level values of coefficient  $d_1$ plotted as  a
function of $1/M_2$ for $\beta=5.9$.  Coefficient $d_1$ equals zero in
the  infinite- and zero-mass  limits and  is typically largest between
charm and bottom.}
\label{fig:d1vsM2}
\end{figure}

Fully $O(a)$-improved    matrix  elements   require    $O(a)$-improved
currents.  Physical fermion  operators, $\Psi$, appear in renormalized
currents.      These     fields   are    constructed     from   fields
$\Psi^{\prime\prime}$, which appear in  the hopping parameter form  of
action $S_f$. Fields $\Psi$ are given by\cite{Kronfeld94}
\begin{eqnarray}
\hat{\Psi} &=
&\left[ 1 + a d_1(m_0, g^2)\,\vec{\gamma}\cdot\vec{D} + O(a^2)\right]
\Psi^{\prime\prime} \label{eqn:Rotation}  \\
\Psi &= &\sqrt{2\kappa}\exp\,(aM_1/2)\hat{\Psi} \FullStop
\label{eqn:NRotation}
\end{eqnarray}
The three-gradient term  is an $O(a)$ correction  to  the current; the
exponential factor is a correction in all-orders of $am_0$.

Coefficient $d_1$  depends upon the  action and  is  a function of the
coupling  and  quark   mass.   \FigRef{fig:d1vsM2} shows  $d_1$  as  a
function of $1/M_2$ for tadpole-improved SW quarks at $\beta=5.9$.  At
tree-level, $d_1$  vanishes in  both zero- and   infinite-mass limits.
The  correction  is typically  largest  between charm  and bottom.  At
$\beta=5.9$, lattice results show  that  $f_D$ increases by  about 7\%
when the three-gradient term is included in the current.

In \EqnRef{eqn:NRotation}   the   factor $\exp(  aM_1/2)$   appears in
addition to the conventional $\sqrt{2\kappa}$ factor for Wilson and SW
quarks.    This factor   removes   large discretization   errors  when
$am_0\simge1$.  Its effect upon Wilson decay constants is seen in
\FigRef{fig:fsqrtM}.   To  $O(am_0)$   this factor   is equivalent, at
tree-level,  to  the  Clover   rotation used   to  improve light-quark
currents.  The  factor becomes $\sqrt{1-6\kappa}$ using the tree-level
expression for  the rest mass   and $\sqrt{1-6u_0\kappa}$ with tadpole
improvement.  The average  plaquette  or kappa critical   is typically
used to estimate $u_0$\cite{LepageMackenzie}.

An $O(a)$-improved current coupling quarks $Q$ and $q$ has the form
\begin{equation}
J^\mu
=\BiLinear{\Psi}{_Q}{\Gamma^\mu}{\Psi}{_q}
=Z_{J}(m_0^Q, m_0^q, g^2)\BiLinear{\hat{\Psi}}{_Q}{\Gamma^\mu}{\hat{\Psi}}{_q}
\label{eqn:Current}
\end{equation}
where the   fields  are   given in  Equations~\ref{eqn:Rotation}   and
\ref{eqn:NRotation}.  Matrix $\Gamma^\mu$   equals  $\gamma^\mu$ for a
vector  current    and $\gamma^\mu\gamma^5$ for   an   axial  current.
Renormalization $Z_{J}$ matches the lattice current to the continuum.

With tadpole improvement, $Z_{J}$ becomes
\begin{equation}
Z_{J}=
\sqrt{2\tilde{\kappa}_Q2\tilde{\kappa}_q}
\exp\left[\frac{1}{2}(a\tilde{M}_1^Q+a\tilde{M}_1^q)\right]\;\tilde{Z}_\Gamma
\label{eqn:Zall}
\end{equation}
where     $\tilde{Z}_\Gamma$    is  the    vertex  correction.   Here,
$\tilde{k}\equiv u_0\kappa$ and $\tilde{M_1}$ denotes tadpole-improved
$M_1$.  $Z_J$  keeps this form to any  order in $g^2$. For example, at
$O(g^2)$                                                    expansions
$\tilde{M}_1=\tilde{M}_1^{[0]}+g^2\tilde{M}_1^{[1]}$               and
$\tilde{Z}_\Gamma=1+g^2 \tilde{Z}_\Gamma^{[1]}$ appear in
\EqnRef{eqn:Zall}. The  higher-order  corrections to  $Z_{J}$ are also
functions of the quark masses.

\begin{figure}
\BorderBox{0pt}{
\epsfxsize=3.0in \epsfbox[82 74 474 316]{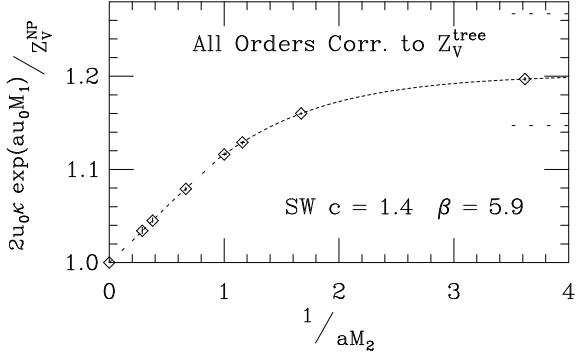}
}
\vskip-1.0cm
\caption{%
All-order  corrections   beyond   tree-level for  the   vector current
renormalization.  Corrections have  a mild dependence upon quark mass.
They  are consistent with  perturbation  theory for  small masses.  At
infinite mass    higher-order     corrections  are  zero    and    the
tadpole-improved tree-level  renormalization  yields the exact result.
}
\label{fig:HigherOrder}
\end{figure}

The  mass dependence of higher-order  terms in $Z_{J}$  can be checked
non-perturbatively for the    vector current by examining   the charge
normalization of    three-point   functions.  \FigRef{fig:HigherOrder}
shows    the   ratio   of   the  tree-level    renormalization factor,
$Z_V^{tree}$,  and the   nonperturbative  factor,  $Z_V^{np}$,  versus
$1/M_2$  for  tadpole-improved SW quarks   at  $\beta=5.9$.  Terms  in
$Z_V^{np}$ beyond  tree level cause this ratio  to differ  from unity.
The figure shows that  higher-order corrections change $Z_V^{tree}$ by
at most  20\% and that light-quark corrections  are largest.  One-loop
corrections   of around   20\%    are expected  for   light  quarks at
$\beta=5.9$.  Note that  renormalization $Z_V^{tree}$ is  exact in the
static limit\cite{BernardLAT93}. Thus, higher-order corrections depend
smoothly on mass  and yield  the expected results  in the  light-  and
infinite-mass limits.

The    Wuppertal   Group\cite{Wuppertal95}   has   also    checked the
renormalization in \EqnRef{eqn:Zall} non-perturbatively.  They look at
ratios   of   the  local-current     three-point function   over   the
conserved-current  three-point function.  They test  \EqnRef{eqn:Zall}
as   function     of quark mass     using  flavor-conserving currents,
heavy-light currents,  and  currents carrying momentum.  The  last two
cases  are   important  tests  of the    current   renormalization for
semileptonic decays.

They  find   that  $Z_V^{tree}$,  given  by \EqnRef{eqn:Zall},   is in
reasonable agreement  with the  nonperturbative determination of $Z_V$
for the three cases  studied.  Because of mass-dependent  higher-order
terms, exact agreement between the nonperturbative renormalization and
$Z_V^{tree}$   is not expected.    In  contrast to  \EqnRef{eqn:Zall},
mass-independent renormalization  using  the  perturbative result  for
light quarks leads to poor  agreement with the nonperturbative results
for heavy quarks.

\subsection{Momentum Dependent Errors}\label{ssect:MomentumErrors}

\newdimen\fs\fs=2.8in
\def\cskp{\hskip12pt}
\def\sskp{\hskip12pt}
\begin{figure*}[t]
\hbox{
\sskp
\BorderBox{0pt}{
\epsfxsize=\fs \epsfbox[96 87 470 242]{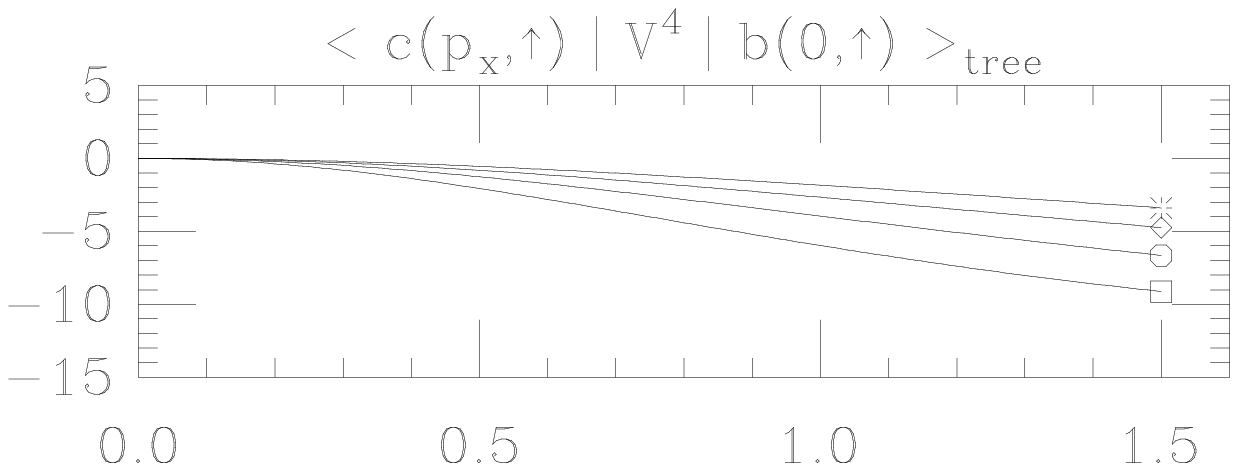}}
\cskp
\BorderBox{0pt}{
\epsfxsize=\fs \epsfbox[96 87 470 242]{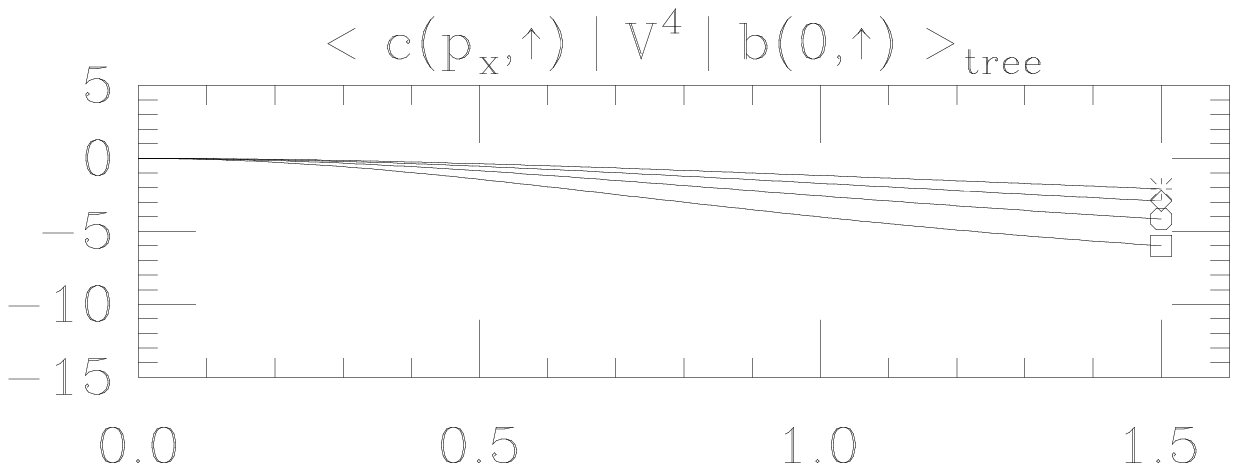}}
}
\hbox{
\sskp
\BorderBox{0pt}{
\epsfxsize=\fs \epsfbox[96 87 470 520]{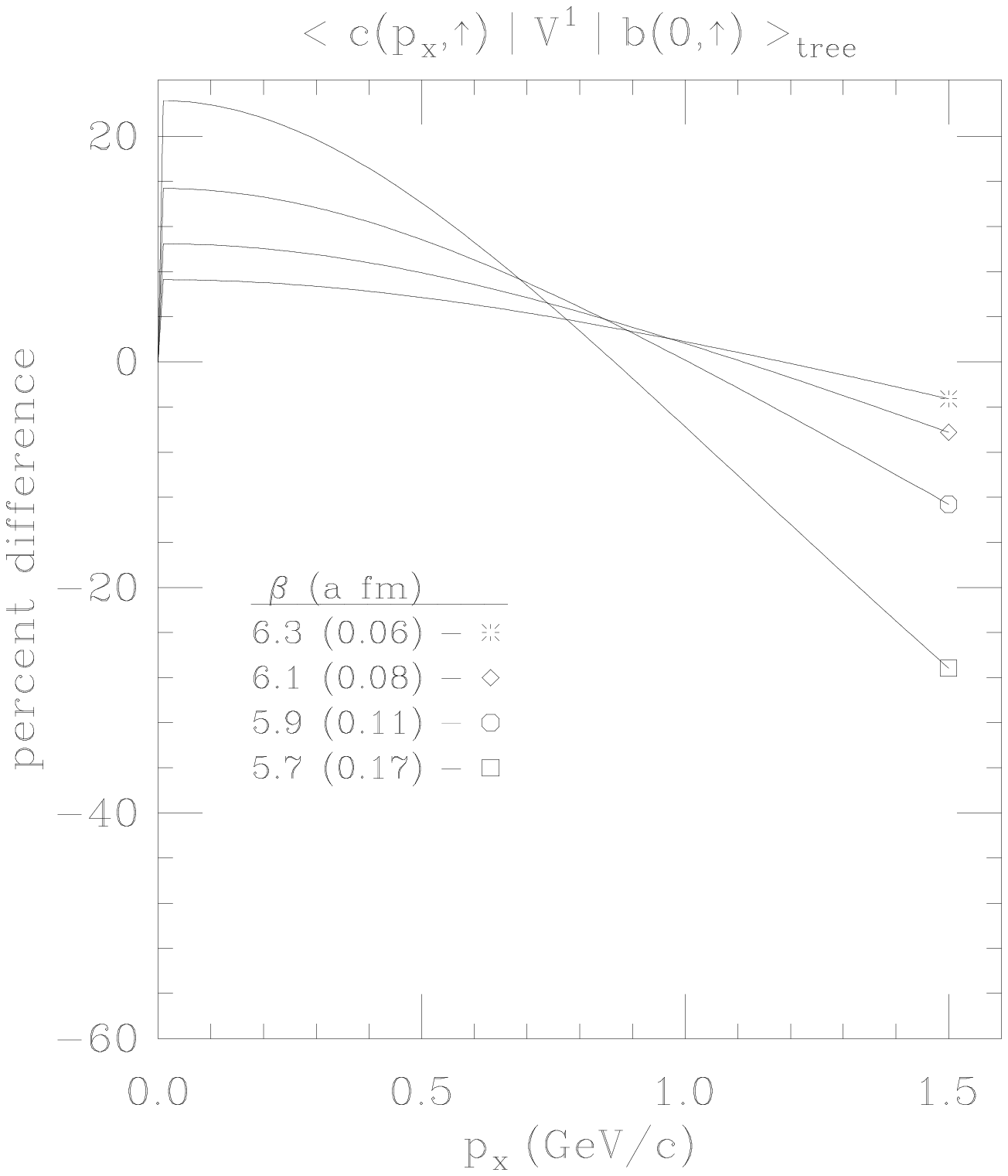}}
\cskp
\BorderBox{0pt}{
\epsfxsize=\fs \epsfbox[96 87 470 520]{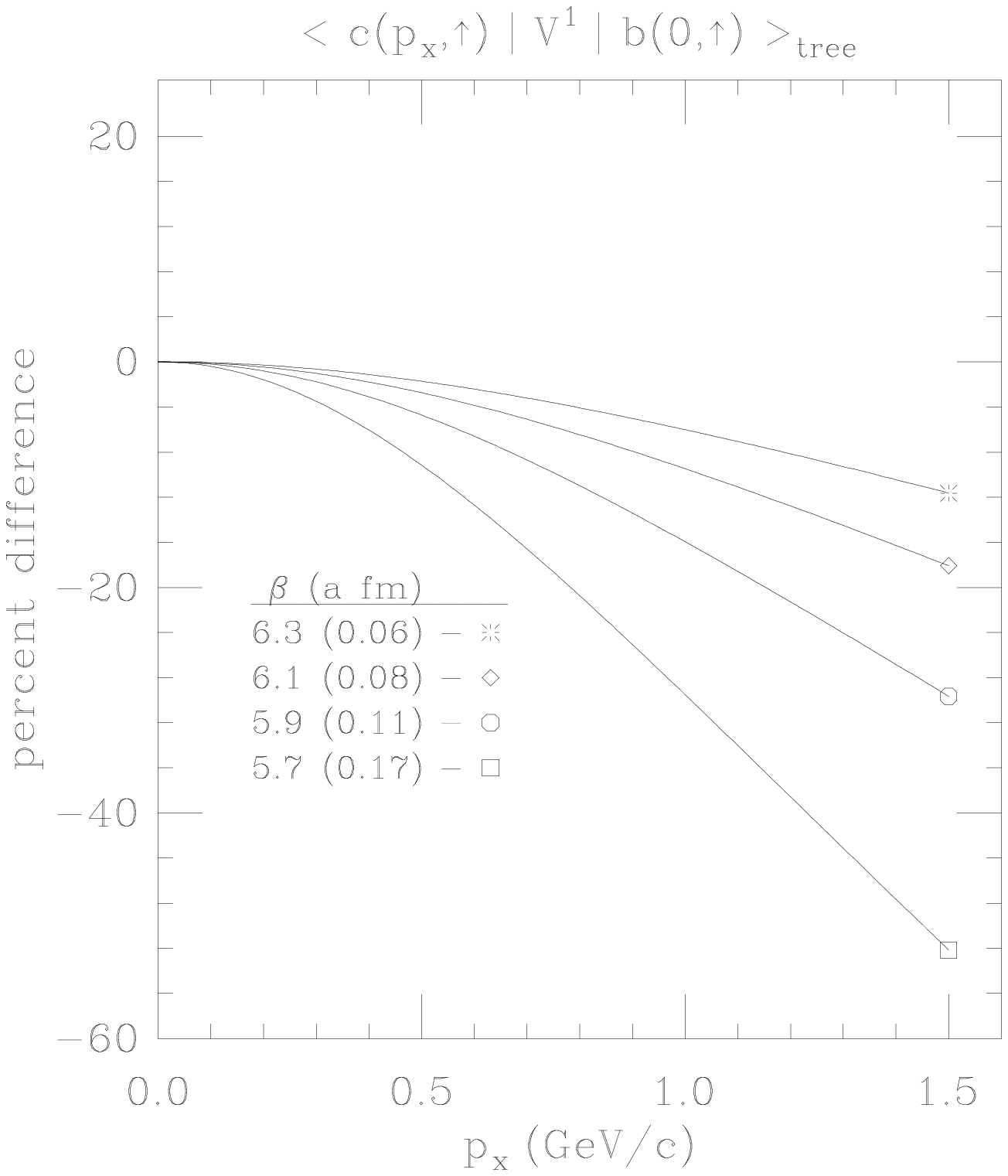}}
}
\label{fig:OapErrors}
\vskip-1.0cm
\caption{%
Momentum   dependent  errors  in   tree-level   quark matrix  elements
$\scriptstyle{\protect\BraKet{c(p_x,\uparrow)}{V^\mu}{b(\vec{0},\uparrow)}}$
for typical lattice spacings.   Errors are shown  as a function of the
charm-quark physical momentum.    Plots show errors in  temporal (top)
and spatial (bottom) matrix elements.  Currents  are the local current
(left) and the $O(a)$-improved current (right).}
\end{figure*}

With fully $O(a)$-improved  actions and currents  $O(a\vec{p})$ errors
are removed from  matrix elements.  Since  improvement  is carried out
for   on-shell    quarks  at    rest,  however,    momentum  dependent
discretization   errors   are  still  a    concern  at  large  momenta
$|a\vec{p}|\gg0$.  It  is  then important to   consider the effect  of
momentum dependent  errors in semileptonic  decays.  Tree-level errors
in quark  matrix elements can  be determined analytically by comparing
infinite-volume lattice matrix elements and continuum matrix elements.
Errors in the quark matrix   elements can then  serve  as a guide   to
estimating errors in hadronic matrix elements.

Consider     matrix elements     for    an  external  vector   current
$\BraKet{q(p_f,\sigma_f)}{V^\mu}  {Q(p_i,\sigma_i)}$ where the current
is defined  in  \EqnRef{eqn:Current}.  Matrix elements for   the local
current ($d_1=0$ in \EqnRef{eqn:Rotation}) and for the $O(a)$-improved
current are calculated.  Wilson quarks with equal kinetic and physical
masses are used.  For simplicity, the initial  quark is at rest.  This
should   be a good approximation to   a heavy quark  decaying within a
meson  at rest.  Since matrix elements  are independent of the mass of
the at-rest quark,  error estimates are the same  for charm and bottom
decays.  Here, the initial quark is taken to be  a bottom quark.  Only
on-axis recoil momenta are considered for the final quark.

\FigRef{fig:OapErrors} shows  the  relative  error in   lattice matrix
elements with a charm-quark final state as  a function of the physical
momentum.  Plots on the left show  errors for the local current, while
those on the right show errors for the $O(a)$-improved current.  Upper
plots correspond  to temporal  matrix elements  while  the lower plots
show spatial matrix elements.  Plots  show typical lattice spacings in
the range $0.17$ to $0.06\,\fm$.

Errors  change more systematically   with  momentum with the  improved
current.  Hence discretization errors  in hadronic matrix elements are
likely to be  under better control at small  momenta with the improved
current.   Note however that  even with full $O(a)$ improvement errors
in spatial  matrix  elements    may   still be    of order  20\%    at
$|\vec{p}|=1.24\,\GeVc$ on  a   $\beta=5.9$ lattice.   A  momentum  of
$1.24\,\GeVc$  corresponds to two units  of lattice  momentum on a two
fermi lattice.  Errors can  be reduced by adding $O(a^2)$ improvements
to the  action and   currents.  Note  however that  precision   matrix
elements     may still be  obtainable    in   the presence of  sizable
discretization errors by extrapolating lattice matrix elements to zero
lattice  spacing at fixed physical  momenta.  The  feasibility of this
approach hinges upon using suitable  techniques to control statistical
noise in lattice correlators with momentum\cite{Onogi94}.

\section{$D$ DECAYS}\label{sect:Ddecays}

Semileptonic decays on the lattice were pioneered  with studies of $D$
mesons\cite{ElKhadra88,Lubicz91}.  $D$ decays continue to be an active
area  of investigation\cite{Lubicz92,Bernard92,Abada94,UKQCD94,APE94}.
At   this  conference     the  Los   Alamos\cite{Bhattacharya94}   and
Wuppertal\cite{Wuppertal95}   groups  presented updates for  their $D$
studies.  Charm decays are an interesting laboratory for QCD and serve
as  a first    step  towards the  study   of $B$   decays.   $D$-decay
experiments provide a    basis  for comparison in   lattice   studies.
Experimental checks for $D$ decays  are an  important way of  checking
systematic errors in lattice calculations.

Experimental  information for $D$  decays has grown steadily since the
first lattice  calculations.  Nevertheless, the  $K$  and $K^*$ recoil
momentum distributions are still not well known\cite{PDG94}.  Instead,
pole dominance  is typically  used to extract  form  factor values  at
maximum recoil    momentum  corresponding to $q^2\equiv(p_D-p_K)^2=0$.
For the $K$, the location of the pole can also  be determined from the
data.  Experiments  find pole  masses consistent with  identifying the
$D_s^*$ as the  dominant pole.  For  $K^*$, poles masses  are fixed to
the  $D_s^*$   and   $D_s^{**}$  masses    and  form   factor   ratios
$R_V=V(0)/A_1(0)$ and   $R_2=A_2(0)/A_1(0)$   are  found.   Individual
vector form  factors are  then found by  assuming the  branching ratio
depends  predominantly  on  $A_1^2$.  Hence  experimental evidence for
pole dominance in $D$ decays is indirect.

Lattice calculations for  $D$ decays cover the  whole range  of recoil
momentum.  Statistical and systematic  errors   at maximum recoil  are
important to estimate since $D$ results are traditionally presented as
form factors  at maximum  recoil.  In the  $D$ rest  frame the maximum
kaon momentum is   $0.86\,\GeVc$ while the  maximum   pion momentum is
$0.93\,\GeVc$. Statistical errors at such  momenta can be minimized by
using good source operators  for  moving mesons. With $\sim300$  gauge
configurations statistical errors below 10\% are  likely for the whole
recoil-momentum range.

Momentum dependent errors  at  maximum  recoil can be  estimated  from
errors in tree-level quark  matrix elements.  Errors in $f^K_+(0)$ are
predicted  to   be   around  10\%  with   $O(a)$-improved   quarks  at
$\beta=5.9$.  This  and other discretization  errors can be reduced by
working on a finer lattice.  With improved actions, however, extremely
fine lattice  spacings are not  necessary.  A  series of  faster  less
costly computations on relatively coarse lattice spacings can be done.
Remaining  discretization errors are then  removed by extrapolation to
zero lattice spacing\cite{Lepage95}.

\begin{table*}[t]
\caption{%
$D   \to K\,l\nu$  and   $D\to K^*\,l\nu$  form  factors  at  $q^2=0$.
Experimental averages  ({\bf  EX}) and a  summary  of  lattice results
({\bf  LA}) are shown.   Lattice spacings, {\bf   a}, lengths, {\bf L}
(both  in $\protect\fm$) and the quark  action ({\bf Ac}) used in each
study are also shown.  }
\label{tab:DtoKsummary}
\begin{tabular}{|l|l|lll||l|l|l|}
\hline
\multicolumn{2}{|c|}{\bf Reference} &{\bf ~~a} &{\bf L} &{\bf Ac} & {\bf f$_+$(0)} & {\bf V(0)} & {\bf A$_1$(0)} \\ \hline \hline
{\bf EX} & Average \cite{PDG94} & & & &$0.75 \pm 0.03$ & $1.1 \pm 0.2$ & $0.56 \pm 0.04$ \\ \hline
{\bf LA} & LMMS \cite{Lubicz92} &.095 &1.0 &W &$0.63 \pm 0.08$ & $0.86 \pm 0.10$ & $0.53 \pm 0.03$ \\
& BKS \cite{Bernard92} &.095 &2.3  &W  &$0.90 \pm 0.08 \pm 0.21$ & $1.43 \pm 0.45 \pm 0.49$ & $0.83 \pm 0.14 \pm0.28$ \\
& ELC \cite{Abada94} &.055 &1.3 &W  &$0.60 \pm 0.15 \pm 0.07 $ & $0.86 \pm 0.24$ & $0.64 \pm 0.16$ \\
& UKQCD \cite{UKQCD94} &.071 &1.7 &SW &$0.67^{+0.07}_{-0.08}$ & $1.01^{+0.30}_{-0.13}$ & $0.70^{+0.07}_{-0.10}$ \\
& LANL \cite{Bhattacharya94} &.095 &3.0 &W &$0.71 \pm 0.04$ & $1.28 \pm 0.07$ & $0.72 \pm 0.03$ \\
& LAT-APE\cite{APE94} &.095 &1.7 &SW &$0.78 \pm 0.08$ & $1.08 \pm 0.22$ & $0.67 \pm 0.11$ \\
& W'tal\cite{Wuppertal95} &.064 &1.5 &W &$0.71\pm0.12^{+0.10}_{-0.07}$ &$1.34\pm0.24^{+0.19}_{-0.14}$ &$0.61\pm0.06^{+0.09}_{-0.07}$ \\ \hline \hline
\multicolumn{2}{|c|}{\bf Reference} &{\bf ~~a} &{\bf L} &{\bf Ac} & {\bf A$_2$(0)} & {\bf V(0)/A$_1$(0)} & {\bf A$_2$(0)/A$_1$(0)} \\ \hline \hline
{\bf EX} & Average \cite{PDG94} & & & &$0.40 \pm 0.08$ & $1.89 \pm 0.25$ & $0.73 \pm 0.15$ \\ \hline
{\bf LA} & LMMS \cite{Lubicz92} &.095 &1.0 &W &$0.19 \pm 0.21$ & $1.6 \pm 0.2$ & $0.4 \pm 0.4$ \\
& BKS \cite{Bernard92} &.095 &2.3  &W  &$0.59 \pm 0.14\pm0.24$ & $1.99 \pm 0.22 \pm 0.33$ & $0.7 \pm 0.16 \pm 0.17$ \\
& ELC \cite{Abada94} &.055 &1.3 &W &$0.40 \pm 0.28 \pm 0.04$ & $1.3 \pm 0.2$ & $0.6 \pm 0.3 $ \\
& UKQCD \cite{UKQCD94} &.071 &1.7 &SW &$0.66^{+0.10}_{-0.15}$ & $1.4^{+0.5}_{-0.2}$ & $0.9 \pm 0.2$ \\
& LANL \cite{Bhattacharya94} &.095 &3.0 &W &$0.49 \pm 0.09$ & $1.78 \pm 0.07$ & $0.68 \pm 0.11$ \\
 & LAT-APE\cite{APE94}&.095 &1.7 &SW &$0.49 \pm 0.34$ & $1.6 \pm 0.3$ & $0.7 \pm 0.4$ \\
& W'tal\cite{Wuppertal95} &.064 &1.5 &W &$0.83\pm0.20^{+0.12}_{-0.08}$ &  & \\ \hline
\end{tabular}
\end{table*}

\TabRef{tab:DtoKsummary} is a summary of $D$   form factors at $q^2=0$
obtained  in  lattice  studies.   Statistical errors  are  shown  with
estimates, when given, for  systematic uncertainties.  Best  estimates
of   form  factors    from    experimental studies    are shown    for
comparison\cite{PDG94}.    To  investigate   lattice-spacing   errors,
lattice spacings, lengths and the quark  action used in each study are
listed in the  table.  Lattice results are mostly  within two sigma of
the experimental values.  Unfortunately,  while statistical errors for
$f_+^K$ are typically below 15\% statistical errors for the other form
factors  tend to be  larger.  Hence, from  the tabulated results it is
difficult to determine how lattice-spacing errors affect form factors.
Higher  precision studies of $D$  decays  are needed.  Comparison with
experiment will be an important  test of whether systematic errors are
under control.

Since discrete lattice  momenta  typically bracket $q^2=0$,   the form
factors in \TabRef{tab:DtoKsummary}  are obtained by interpolation or,
in  cases where statistical   errors were too large, by  extrapolation
over a limited range of  recoil momentum.  In practice, pole dominance
provides a simple means of smoothly interpolating lattice results.

The Los Alamos\cite{Bhattacharya94}   and  Wuppertal\cite{Wuppertal95}
groups  both test possible  interpolation functions  -- including pole
dominance forms.   The Los Alamos group note  that at leading order in
$1/m_c$  the heavy quark symmetry   relation between $D$ form  factors
$f_+$ and  $f_0$  appears to be  incompatible  with both  form factors
having a pole form.  They find $f_+^K$ favors a pole mass lighter than
the   $D_s^*$ mass  while $f_0$ yields   the expected  pole mass.   To
determine  $f_+^K(0)$  they  use  the   pole  form giving the  lighter
effective pole rather than fixing the pole mass to the $D_s^*$ mass.

Although functions inspired  by pole dominance  are commonly used  for
interpolations of lattice $D$ form factors it is important to remember
that how adequately pole dominance  describes $D$ decays or especially
$B$ decays is not known.

\section{$B$ DECAYS TO LIGHT MESONS}\label{sect:BuDecays}

The light-meson decay  modes for   the  $B$ are  especially  important
calculations  for the lattice now  that the CLEO  II collaboration has
branching ratios for the  pion and rho\cite{Gibbons95}.  With a sample
of 2.6 million   $B\;\overline{B}$ decays they  find  branching ratios
(scaled by $\times10^{-4}$)
\def\skpa{\hskip-0.5em}
\def\skpb{\hskip-0.5em}
\def\skpc{\hskip-0.5em}
\def\skpd{\hskip-3.4em}
\begin{eqnarray}
\nonumber
{\cal B}(B^0\to\pi^-\,l^+\nu)&\skpa=&\skpb
\matrix{
1.34\Err(0.35)\Err(0.28)&\skpc{\rm (ISGW)}\cr
1.63\Err(0.46)\Err(0.34)&\skpc{\rm (BSW)}\cr
} \\
\nonumber
{\cal B}(B^0\to\rho^-\,l^+\nu)&\skpa=&\skpb \\
\nonumber
&\skpa &\skpb\skpd\matrix{
2.28\Err(0.36)\Err(0.59)\aErr(+0.00-0.46)&\skpc{\rm (ISGW)}\cr
3.88\Err(0.54)\Err(1.01)\aErr(+0.00-0.78)&\skpc{\rm (BSW)}\cr
} \FullStop
\end{eqnarray}
The  first error for  each  ratio is   statistical and  the second  is
systematic.   The third error for  the rho is  the  uncertainty due to
non-resonant  $\pi\pi$ contributions.   These branching  ratios depend
upon theoretical input for  the efficiency calculation.  The branching
ratios above result when   the ISGW\cite{ISGW} and  the  WSB\cite{BSW}
quark  models  are used  for   efficiencies.  Differences between  the
models indicate that  theoretical uncertainties in the hadronic matrix
elements  are  a substantial  source  of  error.   Note that  reliable
hadronic matrix  elements are   even  more important  for  determining
$|V_{ub}|$.

Light   decay  modes  are     under  investigation  on    the  lattice
\cite{Abada94,APE94,UKQCDBtoPi95,UKQCDBtoRho95,Wuppertal95}.     These
studies  all use   either  standard Wilson  or  Clover\cite{Heatlie91}
quarks.  Because    of concerns  over discretization    errors, direct
calculations involve  heavy mesons with   masses around the  $D$ mass.
Heavy quark symmetry is then used to guide extrapolations from the $D$
system to the $B$ meson.

Hadronic  matrix elements for   light-meson decay  modes scale  in the
heavy-quark  limit\cite{Isgur90,Burdman94}.    For  example,  consider
decays $H\to\pi\,l\nu$. In the $H$-meson rest frame
\begin{equation}
\alpha_s\left(m_H\right)^{2/\beta_0}\;
\frac{\BraKet{\pi(\vec{p}_\pi)}{V^\mu}{H(\vec{0})}}{\sqrt{2m_H\,2E_\pi}}
\sim{\cal M}(p_\pi)
\label{eqn:HQS}
\end{equation}
where ${\cal M}$ is independent of  the heavy-meson mass. The relation
is    not an equality   since there  are   $O(1/m_H)$ corrections.  In
particular, the recoil     momentum     is assumed   to   be     small
$|\vec{p}_\pi|/m_H\ll 1$.  \EqnRef{eqn:HQS} relates $D$ and $B$ matrix
elements with equal pion momentum.  If $O(1/m_H)$ scale-breaking terms
were small this expression  would provide useful information about $B$
decays from experimental data for $D$ decays.

Form factor scaling  relations follow  from matrix element  relations.
For    example,      from       \EqnRef{eqn:HQS}      one        finds
$f_+/\sqrt{m_H}\sim\gamma_+(1+\delta_+/m_H)$ (modulo logarithms) where
the  linear  $1/m_H$  scale-breaking  term  is  now explicit.   On the
lattice, linear functions are typically used for extrapolations.  In a
pioneering study  using this method the  ELC group found $1/m_H$ terms
which  differ  from     zero by  at   most   two  sigma\cite{Abada94}.
Statistical  errors however did     not exclude slopes  as    large as
$1\,\GeV$  for  $f_+/\sqrt{m_H}$ and  $0.7\,\GeV$ for $A_1\sqrt{m_H}$.
These statistical errors and possible systematic errors in heavy-quark
extrapolations are sources of uncertainty in $B$-meson results.

\begin{figure*}[t]
\hbox{
\BorderBox{0pt}{
\epsfxsize=3.0in \epsfbox[93 72 494 403]{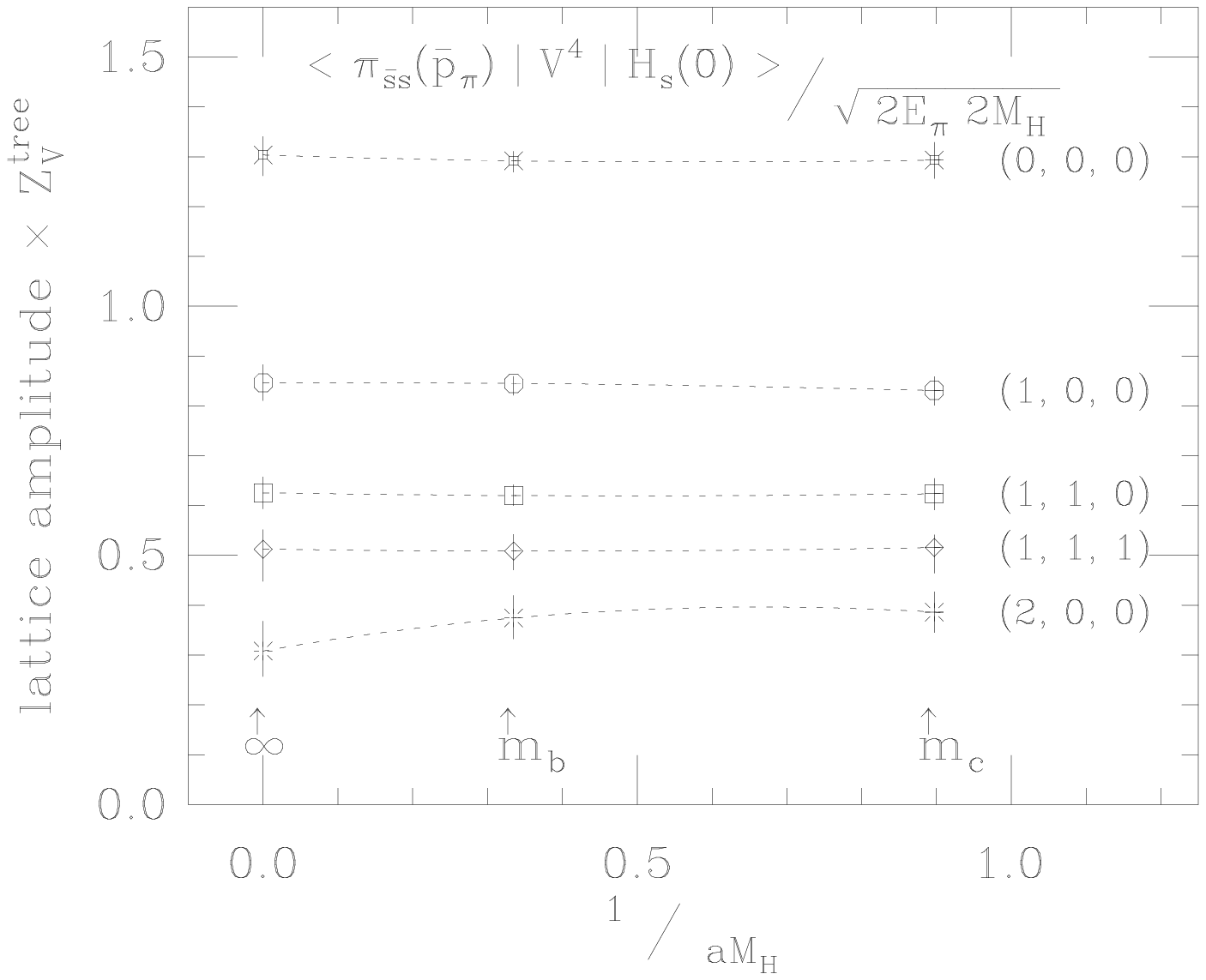}
}
\BorderBox{0pt}{
\epsfxsize=3.0in \epsfbox[93 72 494 403]{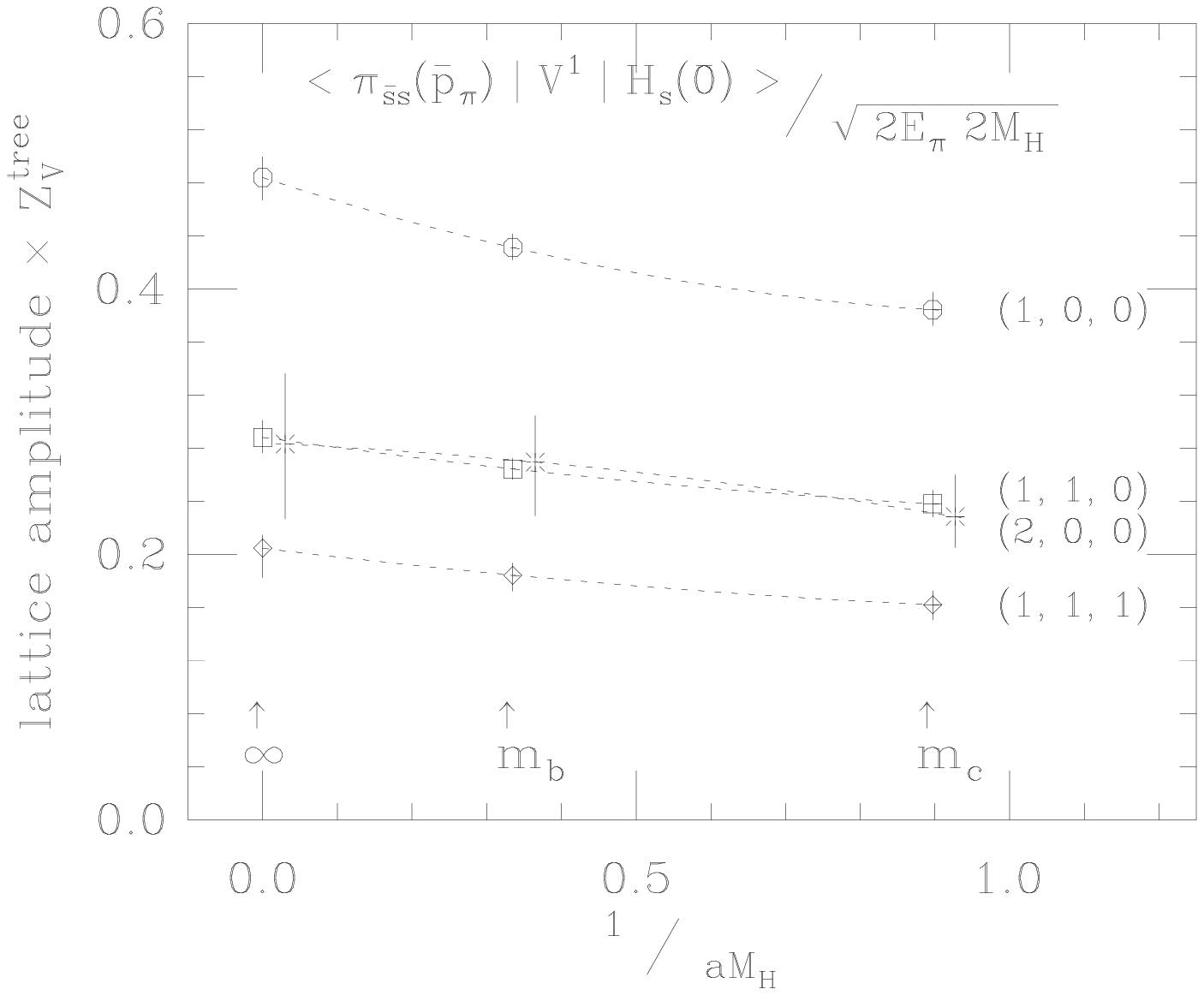}
}
}
\vskip-1.0cm
\caption{%
Hadronic  matrix elements  for  static,  bottom,  and charm quarks  at
$\beta=5.9$. The plot on the left shows temporal  matrix elements.  On
the right are $x$-direction matrix  elements which are proportional to
the $x$-component of  $\vec{p}_\pi$.  Curves and distinct plot symbols
indicate pion momenta which are shown in lattice units.  }
\label{fig:HMEvsM}
\end{figure*}

The       quantity  $\BraKet{\pi}{V^\mu}{H}/\sqrt{2M_H\,2E_\pi}$    in
\EqnRef{eqn:HQS} is obtained directly in lattice calculations when the
heavy meson is at rest.  The dependence upon heavy-quark mass has been
investigated by the  Fermilab group on a  $\beta=5.9$ lattice.  Matrix
elements were computed   using fully $O(a)$-improved charm  and bottom
quarks with strange-mass light quarks.  Results  were also obtained in
the  static approximation.    \FigRef{fig:HMEvsM}  shows  the hadronic
matrix elements  parameterized   by pion momentum  and  plotted versus
$1/m_H$.  Plot  symbols and dashed   curves distinguish pion  momentum
values.  Momenta in lattice units are indicated in the plots; one unit
of momentum corresponds to $0.70\,\GeVc$.  The plot  on the left shows
temporal   matrix elements and on the   right are $x$-component matrix
elements.  Matrix elements  vary  smoothly from  charm  to  the static
limit.  Therefore in studies  employing extrapolations from the $D$ to
the $B$, results are  likely  to be  more reliable when  static matrix
elements  are  included so that    $B$  results are then  obtained  by
interpolation.

Light-meson decay modes for the $B$ have a  maximum recoil momentum of
$2.6\,\GeVc$ compared to only $0.93\,\GeVc$  for $D$ decays.  Concerns
over momentum  dependent discretization  errors and statistical errors
which  increase rapidly with   momentum limit current calculations  to
momenta around $1\,\GeVc$.  Hence in contrast to  $D$ decays, $B$ form
factors at $q^2=0$ are far beyond the range of direct calculation.  In
addition, extrapolations to large recoil momenta are also likely to be
unreliable.  Extrapolations    to large  momenta   can amplify   small
discretization errors  which      occur for small    momenta.    Also,
extrapolations  which assume pole dominance  for  the $B$ introduce an
unknown degree  of model  dependence.   Therefore it is  best to focus
upon quantities in $B$ decays  that can be directly calculated without
extrapolations.

\begin{figure}
\BorderBox{0pt}{
\epsfxsize=3.0in \epsfbox[96 86 474 374]{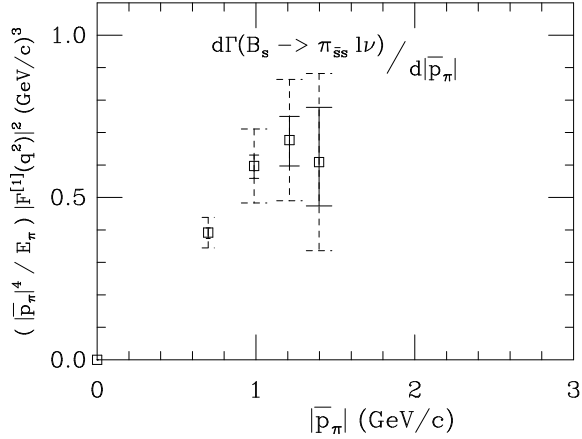}
}
\vskip-1.0cm
\caption{%
$B$-meson   differential   decay  width   for  the  light-pseudoscalar
mode. Light-quark masses are approximately equal  to the strange quark
mass.  Solid error   bars are statistical errors.  Estimated  momentum
dependent errors are shown by the larger broken-line error bars.}
\label{fig:GammaBtopi}
\end{figure}

Differential  decay   rates and partial widths    do not require large
extrapolations in momentum.   Consider the differential decay rate for
$B\to\pi\,l\nu$
\begin{equation}
\frac{d\Gamma}{d|\vec{p}_\pi|}=
2m_B\frac{G_f^2|V_{ub}|^2}{24\pi^3}\,
\frac{|\vec{p}_\pi|^4}{E_\pi}\;|\,f_+^\pi(q^2)\,|^2 \FullStop
\label{eqn:DiffRate}
\end{equation}
\FigRef{fig:GammaBtopi}   shows this   differential  rate  (less the
momentum independent pre-factors) in a preliminary study of $B$ decays
by the Fermilab group.  Statistical errors are shown by the solid-line
error bars;  larger (broken-line)  error bars show  momentum dependent
errors estimated using tree-level    quark matrix elements.    In this
$\beta=5.9$  study   statistical errors are  below   10\% and momentum
dependent errors are estimated to be less  than 15\% below $1\,\GeVc$.
Errors may then  be small enough  that extrapolations to  zero lattice
spacing are reliable enough to remove most discretization errors.

Results in  \FigRef{fig:GammaBtopi} are for strange-mass light quarks.
Chiral  extrapolations  are necessary to   obtain decay rates  for the
physical  pion.  Chiral extrapolations  of  form factor $f_+$ at small
pion momenta may be difficult  since soft pion arguments suggest $f_+$
varies rapidly with  pion mass in  heavy-meson decays\cite{Burdman94}.
Note   however variations   in $f_+$ at     small recoil momenta   are
suppressed by four  powers of the momentum in  the decay rate.  Hence,
chiral extrapolations  of differential decay rates  may be more robust
than chiral extrapolations of form factors.

\begin{table}
\caption{%
$B$-meson  differential rates and  partial widths for the light-vector
decay mode.   Momenta  are in  $\protect\GeVc$.  Width  $\Gamma$ is in
units         of     $|V_{ub}|^2\protect\times10^{-12}\,\protect\GeV$.
Strange-mass           light     quarks      are      used.      UKQCD
collaboration\protect\cite{UKQCDBtoRho95}.}
\label{tab:UKQCDBtoRho}
\renewcommand{\arraystretch}{1.2}
\begin{tabular}{lll}
\hline
$q^2_0$			&$d\Gamma/dq^2|_{q^2_0}$ &$\int_{q^2_0}^{q^2_{max}}\,d\Gamma$ \\
\hline
$20.3$			&$0.00$			&$0.0$ \\
$19.7\aErr(+1-1)$	&$0.19\aErr(+3-2)$	&$0.08\aErr(+1-1)$ \\
$17.5\aErr(+2-2)$	&$0.57\aErr(+6-5)$	&$0.9\aErr(+1-1)$ \\
$16.7\aErr(+2-2)$	&$0.58\aErr(+9-6)$	&$1.3\aErr(+1-1)$ \\
$15.3\aErr(+3-3)$	&$0.6\aErr(+1-1)$	&$2.3\aErr(+2-2)$ \\
$14.4\aErr(+3-3)$	&$0.8\aErr(+2-1)$	&$3.0\aErr(+3-2)$ \\
\hline
\multicolumn{3}{l}{charm threshold} \\
$11.6$			&---			&$5.4\aErr(+7-5)$ \\
\hline
\end{tabular}
\end{table}

Partial   widths can be  obtained  by  smoothly interpolating  between
lattice differential decay rates values and then integrating.  Partial
widths obtained    this way are  expected    to  be  relatively  model
independent   for momenta   within   the  range   of  direct   lattice
calculations.   The UKQCD  collaboration  has found widths  using this
method              for        the             vector            decay
mode\cite{UKQCDBtoRho95}. \TabRef{tab:UKQCDBtoRho} shows their results
for differential decay rates and partial widths along with statistical
errors.  Chiral   extrapolations have not  been done  and strange-mass
light  quarks are  used.    Note that at   $q^2_0=14.4(\GeVc)^2$ which
corresponds   to  $|\vec{p}_\rho|=1.2\,\GeVc$  statistical errors  are
about 10\% for  the partial width. The authors  point out that partial
widths for the rho and  pion decay modes  from the lattice can be used
to  extract $|V_{ub}|$ in a  model  independent way from exclusive $B$
decay modes.

The  CLEO II exclusive-mode analysis suggests  that a sizable fraction
of  pions have  momenta  below $1.4\,\GeVc$.   Hence exclusive partial
widths from the lattice  may in fact  be a viable means  of extracting
$|V_{ub}|$.   Minimizing  errors in $|V_{ub}|$ will  require adjusting
the maximum recoil   momentum limit  in   the partial width so  as  to
minimize both lattice and experimental errors.

The most important task for   semileptonic decay studies is  improving
the  precision to which $|V_{cb}|$ is  known.  It is then important to
eliminate potential sources  of error.  Extrapolations to large recoil
momenta using  present lattice  results are unreliable.   Differential
decay  rates and partial  widths are  likely  to be more  reliable and
present  lattice results in  a form  more useful  to the  experimental
community.

\section{DECAYS TO HEAVY HADRONS}\label{sect:BcDecays}

Heavy-meson  decay modes are studied  to test non-perturbatively heavy
quark  symmetry and  to provide  theoretical  input needed  to extract
$|V_{cb}|$ from $B$-meson decay rates.  The study of these decay modes
on the lattice is discussed in detail  by Kenway\cite{Kenway93} and by
Lellouch\cite{Lellouch94}. This section provides a  update in the form
of results presented at this conference.

The  Isgur-Wise  function  (up   to  $O(1/m)$   corrections) has  been
extracted  from QCD matrix  elements  calculated using  Wilson and  SW
quarks\cite{Shen93,UKQCDBtoD}.  Systematic   error estimates for these
calculations are not  yet  complete. Tree-level quark  matrix elements
however provide  a way of  estimating momentum-dependent errors.  Note
that according to \FigRef{fig:OapErrors}  relative errors  are smaller
for temporal  matrix    elements.  Hence elastic   form factors  which
require only temporal matrix elements are likely to yield the smallest
momentum-dependent uncertainties  for  the  Isgur-Wise function.   The
figure shows  that tree-level errors  are probably  less than 10\% for
momenta below $1\,\GeVc$.   It would be  interesting to see how errors
bias the determination  of  the Isgur-Wise  slope at  the  zero-recoil
point.

Another means of  studying heavy quark symmetry  is to formulate Heavy
Quark Effective Theory (HQET) directly on the lattice\cite{Mandula93}.
At leading order lattice HQET (LHQET)  describes an infinitely massive
quark  with arbitrary  four-velocity.     The lattice theory must   be
renormalized to match HQET.

At this   conference,   Mandula   and  Ogilvie   investigate  velocity
renormalization\cite{Mandula95}.    In     HQET   the   multiplicative
renormalization of the    heavy quark's four-velocity   is  actually a
wavefunction renormalization.    Lorentz symmetry  is reduced   on the
lattice  to hypercubic symmetry.  Then, in  LHQET additive shifts  can
also   occur.      These   renormalizations    have    been   computed
perturbatively\cite{Aglietti}.   The   perturbative   expectations are
compared with nonperturbative  determinations of  the  renormalization
coefficients from  the lattice.  In  this pilot study, the coefficient
of the  term proportional to four-velocity  is within one sigma of the
one-loop value.  Perturbative and lattice coefficients for terms cubic
in the four-velocity were  in complete disagreement  however.  Tadpole
improvement of  the LHQET action  and a Lepage-Mackenzie  treatment of
perturbation   theory\cite{LepageMackenzie} may   help  resolve  these
discrepancies.

The Kentucky  group  presented preliminary numerical  results  for the
unrenormalized Isgur-Wise form  factor  in LHQET\cite{Draper95}.  They
find   their  variational   method  (MOST) for    constructing optimal
wavefunctions  can be effectively used  in LHQET.  MOST operators have
good signal-to-noise ratios.  With 32 gauge configurations statistical
errors are around 5\% while errors due to excited state contaminations
are estimated to be 5\%.   These preliminary results encourage further
study. Comparisons of the Isgur-Wise  function directly computed using
LHQET and the  Isgur-Wise function extracted using finite-mass  quarks
will provide information about $O(1/m)$ errors.

Baryons containing  one heavy   quark are  under   study by the  UKQCD
collaboration\cite{Richards95}.  Semileptonic  decays of baryons   may
provide additional checks on CKM  matrix elements.  Baryons also offer
additional tests of  heavy  quark  symmetry.   Consider  the six  form
factors for  $\Lambda_b\to\Lambda_c\,l\nu$.   They are   related to  a
single baryon Isgur-Wise function appearing  in the heavy quark limit.
The UKQCD  collaboration presented  preliminary results for  the axial
form  factor  which is  expected to have  small  $O(1/m)$ corrections.
They find little dependence  upon $1/m$ when  the heavy-quark  mass is
varied about the charm-quark mass.

Heavy hadron decays are important for determining $|V_{cb}|$.  Lattice
calculations provide   valuable nonperturbative  tests of heavy  quark
symmetry as well  as Isgur-Wise form factors.   HQET on the lattice is
still being explored.    In LHQET calculations  statistical noise  and
poor ground-state  isolation are  important problems  to  be overcome.
Highly optimal particle operators are then necessary.

\section{REMARKS}\label{sect:Remarks}

Semileptonic  decays provide  important information   about CKM matrix
elements  involving heavy quarks.   Recent  CLEO  II results  for  the
light-meson decay  modes  of the $B$ are   exciting since they promise
better  determinations of $|V_{ub}|$.   Since reliable matrix elements
are a  crucial to  this  goal, $B$  decays  deserve  high priority  in
lattice studies.

Systematic errors must be  understood to obtain reliable  results from
the   lattice.  Discretization   errors  are  an   important source of
uncertainty.  Improvement  does not  require expensive  lattices  with
drastically  smaller  lattice   spacings  to decrease   discretization
errors.  Improvement is  possible  with arbitrary-mass  four-component
quarks.  Bottom quarks can   then    be put on  existing     lattices.
Extrapolations in the heavy-quark mass which are a potential source of
systematic error in bottom-quark studies can then be eliminated.

With $O(a)$ improvement  momentum dependent  discretization errors  in
matrix elements  are still a concern on   present lattices for momenta
above $1\,\GeVc$.   Alternatives to  drastically reducing the  lattice
spacing as   a way of  controlling  momentum dependent errors include:
$O(a^2)$ improvement, and  extrapolations to  zero lattice spacing  at
fixed physical momenta.

Present  $D$  decay   calculations cover  the   full   range of recoil
momentum.    Careful attention to   sources of systematic error should
lead to reliable matrix  elements.   Experimental data for $D$  decays
provide a basis for comparing lattice  results. Checks for the $D$ are
important for testing that systematic errors are under control.

For light-meson decay modes of the $B$ the  maximum recoil momentum is
greater than the largest momentum than can be confidently studied with
present   lattices.    Attention  should  then   be   directed towards
quantities that can be  reliably determined.  Differential decay rates
and partial widths are  a useful form for  lattice results that do not
require extrapolations to  large recoil momenta.  Partial widths  from
the lattice may offer a viable means of extracting $|V_{ub}|$ from the
CLEO II exclusive decay measurements.

\section*{ACKNOWLEDGMENTS}

I thank T.\  Bhattacharya, T.\ Draper,  J.\ Flynn, C.\ McNeile and D.\
Richards for discussions and  information  about their work.   I thank
L.\  Gibbons and  E.\  Thorndike for   discussions concerning  CLEO II
exclusive  decay results.  I  wish to thank  my colleagues: B.\ Gough,
G.\ Hockney, A.\  El-Khadra, A.\ Kronfeld, B.\  Mertens, T.\ Onogi and
P.\  Mackenzie for     an interesting  and   productive collaboration.
Informative   discussions   with members  of  the   FNAL  theory group
especially W.\ Bardeen, G.\  Burdman and E.\ Eichten are acknowledged.
Fermilab  calculations are  performed  on the  ACPMAPS  supercomputer.
Fermilab is  operated by University Research  Association, Inc.  under
contract with the U.S. Department of Energy.

\end{document}